\documentclass[sigconf,nonacm]{acmart}
\AtBeginDocument{%
  }

\usepackage{multirow}
\usepackage{enumitem}
\usepackage{subcaption}
\usepackage{listings}
\usepackage[most]{tcolorbox}
\usepackage{microtype}
\usepackage{xspace}
\newcommand{\aiopsds}{\textit{AIOps2025}\xspace}
\newcommand{\rcads}{\textit{RCA100}\xspace}

\definecolor{gtkey}{RGB}{125,80,160}    
\definecolor{gtstr}{RGB}{40,120,40}     
\definecolor{gtnum}{RGB}{170,90,30}     
\definecolor{gtcom}{RGB}{120,120,120}   
\lstdefinelanguage{gtjson}{
  morestring=[b]",
  morecomment=[l]{//},
  basicstyle=\ttfamily\footnotesize,
  showstringspaces=false,
  keywordstyle=\color{gtkey}\bfseries,
  stringstyle=\color{gtstr},
  commentstyle=\color{gtcom}\itshape,
  morekeywords={true,false,null},
  breaklines=true,
  breakatwhitespace=true,
  breakindent=2em,
  literate=
    *{:}{{{\color{gtkey}:}}}{1}
     {0}{{{\color{gtnum}0}}}{1}
     {1}{{{\color{gtnum}1}}}{1}
     {2}{{{\color{gtnum}2}}}{1}
     {3}{{{\color{gtnum}3}}}{1}
     {4}{{{\color{gtnum}4}}}{1}
     {5}{{{\color{gtnum}5}}}{1}
     {6}{{{\color{gtnum}6}}}{1}
     {7}{{{\color{gtnum}7}}}{1}
     {8}{{{\color{gtnum}8}}}{1}
     {9}{{{\color{gtnum}9}}}{1},
}
\newtcolorbox{gtbox}[1][]{%
  enhanced, rounded corners=all,
  colback=yellow!8, colframe=black!50,
  boxrule=0.4pt, arc=2mm, left=4pt, right=4pt, top=4pt, bottom=4pt,
  fontupper=\footnotesize, #1}

\setcopyright{none}
\copyrightyear{2026}
\acmYear{2026}

\begin{document}

\title{A Multi-Dataset Benchmark for Evaluating LLM Agents in Microservice Failure Diagnosis}

\author{Yuanhong Cai}
\affiliation{%
  \institution{Computer Network Information Center, Chinese Academy of Sciences}
  \city{Beijing}
  \country{China}}

\author{Xiaohui Nie}
\affiliation{%
  \institution{Computer Network Information Center, Chinese Academy of Sciences}
  \city{Beijing}
  \country{China}}

\author{Kanglin Yin}
\affiliation{%
  \institution{Key Laboratory for Satellite Digitalization Technology,
    Chinese Academy of Sciences}
  \city{Shanghai}
  \country{China}}

\author{Changhua Pei}
\affiliation{%
  \institution{Computer Network Information Center, Chinese Academy of Sciences}
  \city{Beijing}
  \country{China}}

\author{Yongqian Sun}
\affiliation{%
  \institution{Nankai University}
  \city{Tianjin}
  \country{China}}

\author{Shenglin Zhang}
\affiliation{%
  \institution{Nankai University}
  \city{Tianjin}
  \country{China}}

\author{Haibin Liu}
\affiliation{%
  \institution{Alibaba Cloud Computing Company}
  \city{Hangzhou}
  \country{China}}

\author{Guiyang Liu}
\affiliation{%
  \institution{Alibaba Cloud Computing Company}
  \city{Hangzhou}
  \country{China}}

\author{Xidao Wen}
\affiliation{%
  \institution{Alibaba Cloud Computing Company}
  \city{Hangzhou}
  \country{China}}

\author{Fang Situ}
\affiliation{%
  \institution{Alibaba Cloud Computing Company}
  \city{Hangzhou}
  \country{China}}

\author{Dan Pei}
\affiliation{%
  \institution{Tsinghua University}
  \city{Beijing}
  \country{China}}

\renewcommand{\shortauthors}{Cai et al.}

\begin{abstract}

LLM-based agents are reshaping microservice operations into \emph{AgentOps}, where benchmarks are key to evaluating failure diagnosis over multimodal observability data.
However, existing benchmarks remain largely outcome-oriented: they score only the final answer and fail to assess the systematic reasoning process in failure diagnosis.
We address this gap by introducing two large-scale datasets (\textbf{\aiopsds} and \textbf{\rcads}) under a reasoning-process evaluation paradigm that assesses agentic diagnostic capability along three dimensions: \emph{Localization}---where the fault occurs, \emph{Identification}---what type of fault it is, and \emph{Reason}---whether the reasoning trace is grounded in relevant evidence. 
Together, the two datasets comprise over $500$ expert-labeled failure cases across two representative microservice systems (HipsterShop and the OpenTelemetry Demo Store). They cover diverse fault scenarios across resource, network, runtime, middleware/database, and application-logic categories and provide fine-grained causal evidence to support agent learning and reasoning-process evaluation. Beyond scale and coverage, the datasets have been carefully labelled by domain experts and validated through large-scale competitions, supporting more than $6{,}000$ participating teams. This makes them not only expert-labeled diagnostic datasets, but also competition-validated benchmarks for evaluating agentic failure diagnosis in real-world microservice environments. Datasets are available at \url{https://www.aiops.cn/gitlab/aiops-live-benchmark/agenticopseval}.
\end{abstract}

\settopmatter{printacmref=false}

\maketitle

\begin{figure}[t]
  \centering
  \includegraphics[width=\columnwidth]{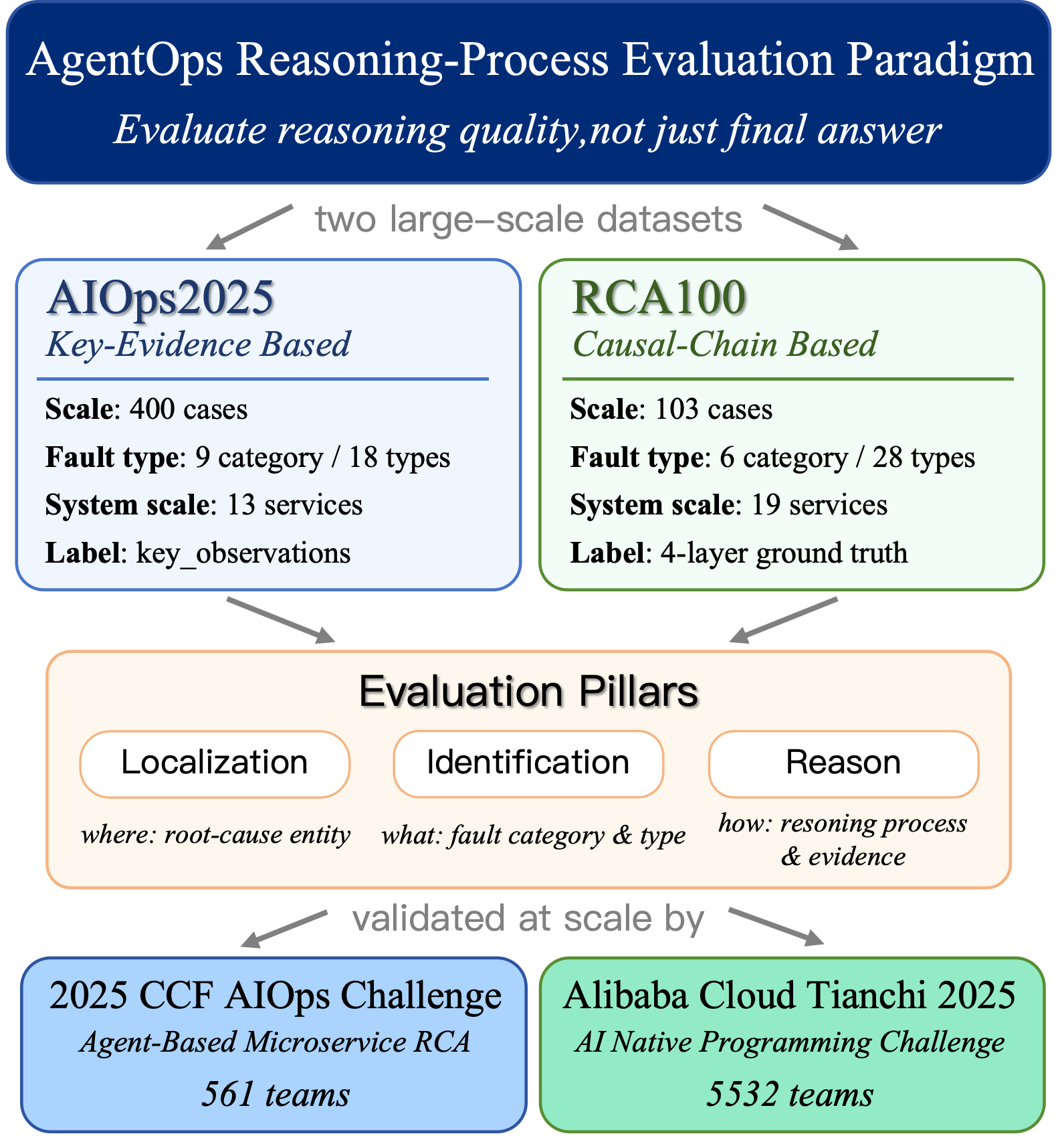}
  \caption{Overview of our benchmark.}
  \Description{Overview figure of the benchmark covering Localization,
  Identification, and Reason pillars instantiated on AIOps2025 and
  RCA100.}
  \label{fig:overview}
\end{figure}

\section{Introduction}
\label{sec:intro}


Microservice reliability is a fundamental concern in modern software operations~\cite{sre-book, failuresurvey}. With chain-of-thought prompting~\cite{cot} and ReAct-style tool use~\cite{react, toolllm}, LLM-based AIOps agents have shown growing capability in diagnosing failures from multimodal observability data, including metrics, logs, and traces~\cite{opentelemetry, prometheus, jaeger}. These advances have shaped the emerging \emph{AgentOps} paradigm, spanning single-agent tool augmentation~\cite{dbot, rcacopilot} and multi-agent collaboration~\cite{sun2025trioxpert, mabc, flowofaction}.

Evaluating agentic diagnostic capability is essential for assessing whether agent systems are practically usable in microservice operations. However, existing benchmarks for root cause analysis~\cite{rcaeval} and LLM-based agents~\cite{opseval} mostly focus on final-answer matching, ignoring how the diagnosis is derived. Such outcome-only evaluation can misjudge agent capability, since a correct answer may result from keyword matching rather than systematic, evidence-grounded reasoning. A more reliable benchmark should therefore evaluate not only root-cause accuracy, but also whether the agent can localize the fault, identify its root cause, and justify the diagnosis with relevant evidence or causal chain. Constructing such a reasoning-process benchmark raises two key challenges:

\begin{itemize}[leftmargin=1.2em,itemsep=0.3em]
\item \textbf{Insufficient reasoning-process evaluation.}
Existing benchmarks typically label only the final root cause, making it difficult to distinguish evidence-grounded diagnosis from accidental keyword matching. A fair benchmark should further specify \emph{which evidence is diagnostically relevant} and \emph{how such evidence supports the causal reasoning process}.

\item \textbf{Lack of large-scale validation.}
Existing datasets are typically tested by only small expert groups. A robust benchmark should also be validated by large-scale users to ensure reliable evaluation across diverse agents.
\end{itemize}


To address these challenges, we present two large-scale microservice AIOps datasets, validated through two major national-level public competitions in 2025, for evaluating agentic anomaly detection, fault localization, and reasoning processes (Fig.~\ref{fig:overview}). We organize the evaluation around three shared pillars: \emph{Localization}---which entity causes the failure, \emph{Identification}---what fault type occurs, and \emph{Reason}---whether the agent grounds its reasoning in relevant evidence. \textbf{\aiopsds} evaluates key-evidence coverage by labelling per-modality key observations for open-ended anomaly descriptions in a self-hosted HipsterShop system. \textbf{\rcads} evaluates causal-chain coverage by providing structured alert inputs and four-layer ground truth---fault category, root-cause entity, causal propagation chain, and evidence checkpoints---on the OpenTelemetry Demo Store deployed on Alibaba Cloud ACK. In summary, the main contributions are as follows.


\begin{itemize}[leftmargin=1.2em,itemsep=0.3em]
\item \textbf{A reasoning-process evaluation paradigm.}
We argue that RCA agent evaluation should move beyond final-answer or component-only scoring and assess whether an agent's reasoning trace is grounded in the right diagnostic evidence. We operationalize this idea through three shared pillars---\emph{Localization}, \emph{Identification}, and \emph{Reason}---instantiated by two complementary labelling forms: \emph{key-evidence} and \emph{causal-chain}.
\item \textbf{Two large-scale multimodal datasets.}
We release $503$ expert-labeled failure cases with $\approx 15.3$\,GB of multimodal observability data from two heterogeneous microservice architectures. Each dataset is paired with a deterministic scoring protocol aligned with the three pillars, turning reasoning-process evaluation into an executable benchmark.

\item \textbf{Large-scale real-world competition validation.}
The datasets powered two national-level public competitions: the \textit{2025 CCF AIOps Challenge}\footnote{\url{https://challenge.aiops.cn/home/competition/1920410697896845344}} ($561$ teams) and \textit{the Alibaba Cloud Tianchi 2025 AI-Native Programming Challenge}\footnote{\url{https://tianchi.aliyun.com/competition/entrance/532387}} ($5{,}532$ teams), involving $6{,}093$ teams in total. To the best of our knowledge, this is the largest competition-validated dataset for microservice reasoning-process evaluation to date. The datasets are publicly available at \url{https://www.aiops.cn/gitlab/aiops-live-benchmark/agenticopseval}.
\end{itemize}

The remainder of this paper is organized as follows. Section~\ref{sec:related} reviews related work, and Section~\ref{sec:design} presents our design principles. Sections~\ref{sec:dataset_a} and~\ref{sec:dataset_b} introduce the two datasets, including their statistical properties and large-scale real-world validation. Section~\ref{sec:lessons} discusses lessons, limitations, and broader uses, before Section~\ref{sec:conclusion} concludes.

\section{Related Work}
\label{sec:related}

Existing failure-diagnosis datasets fall into three categories:
those for traditional RCA methods, those for LLM-based agents,
and interactive environments for agent training.
Table~\ref{tab:bench-comparison} contrasts them with our work on
case scale, modality coverage, reasoning-process labelling, and
large-scale validation.

\begin{table*}[!t]
\centering
\caption{Existing AIOps benchmarks vs.\ our work on case scale,
modality coverage, reasoning-process labelling, and large-scale
validation. M / L / T / E / A / Topo = Metrics / Logs / Traces /
Events / Alerts / Topology. $\checkmark$ = supported, $\times$ =
not supported. ``Partial'' for ITBench means the ground truth
includes the fault propagation chain and remediation steps, but
not the per-step diagnostic evidence checkpoints.}
\label{tab:bench-comparison}
\small
\setlength{\tabcolsep}{4pt}
\begin{tabular}{lrllc}
\toprule
Benchmark & \#Cases & Modalities & Reasoning process label & Large-scale validation \\
\midrule
RCAEval~\cite{rcaeval}             & $735$       & M$+$L$+$T & $\times$                                 & $\times$ \\
OpenRCA~\cite{openrca}             & $335$       & M$+$L$+$T & $\times$                                 & $\times$ \\
AIOpsLab~\cite{aiopslab}           & $88$        & M$+$L$+$T & $\times$                                 & $\times$ \\
ITBench~\cite{itbench}             & $94$        & M$+$L$+$T & $\checkmark$ (partial)                   & $\times$ \\
SREGym~\cite{sregym}               & $114$       & M$+$L$+$T & $\times$                                 & $\times$ \\
\midrule
\textbf{\aiopsds (ours)}          & $\mathbf{400}$ & M$+$L$+$T & $\checkmark$ (key-evidence)         & $\checkmark$ (CCF / $561$ teams) \\
\textbf{\rcads (ours)}          & $\mathbf{103}$ & M$+$L$+$T$+$E$+$A$+$Topo & $\checkmark$ (causal chain) & $\checkmark$ (Tianchi / $5{,}532$ teams) \\
\bottomrule
\end{tabular}
\end{table*}

\textbf{Datasets for traditional RCA.} RCAEval~\cite{rcaeval}
aggregates $735$ cases across three open-source microservice
systems with component-level root-cause labels, serving as the
de-facto baseline for causal-graph and change-point methods
(e.g., MicroScope~\cite{microscope}, MicroRCA~\cite{microrca},
RCD~\cite{rcd}, BARO~\cite{baro}, DiagFusion~\cite{diagfusion}).
The labels record \emph{what} the answer is, not \emph{how} a
reasoner should reach it, leaving evidence-grounded diagnosis
indistinguishable from keyword luck.

\textbf{Datasets for LLM-based agents.} OpenRCA~\cite{openrca} is
the representative offline benchmark, asking an LLM to output a
$\langle$time, component, reason$\rangle$ triple over $335$ cases
from three enterprise systems. It still grades the final answer
only. Our two datasets enter this category and add explicit
reasoning-process supervision: per-modality key evidence
(\aiopsds) and a causal propagation chain with $661$ evidence
checkpoints (\rcads), so scores reflect diagnostic process
quality.

\textbf{Interactive environments for agent training.}
AIOpsLab~\cite{aiopslab} ($88$ tasks), ITBench~\cite{itbench}
($94$ scenarios), and SREGym~\cite{sregym} ($114$ problems) put
agents inside live systems and score end-to-end execution.
Evaluation is dominantly pass/fail or efficiency-only; only
ITBench partially labels the reasoning process. These environments
are complementary deployment targets, not substitutes: our datasets
supply the missing process-level supervision and the first
large-scale competition validation.

\section{Design Principles}
\label{sec:design}

A useful RCA benchmark for LLM agents must control three axes:
input-signal richness, fault-coverage breadth, and observability
of the reasoning process. We adopt one design principle along
each.

\begin{figure*}[t]
  \centering
  \includegraphics[width=0.95\textwidth]{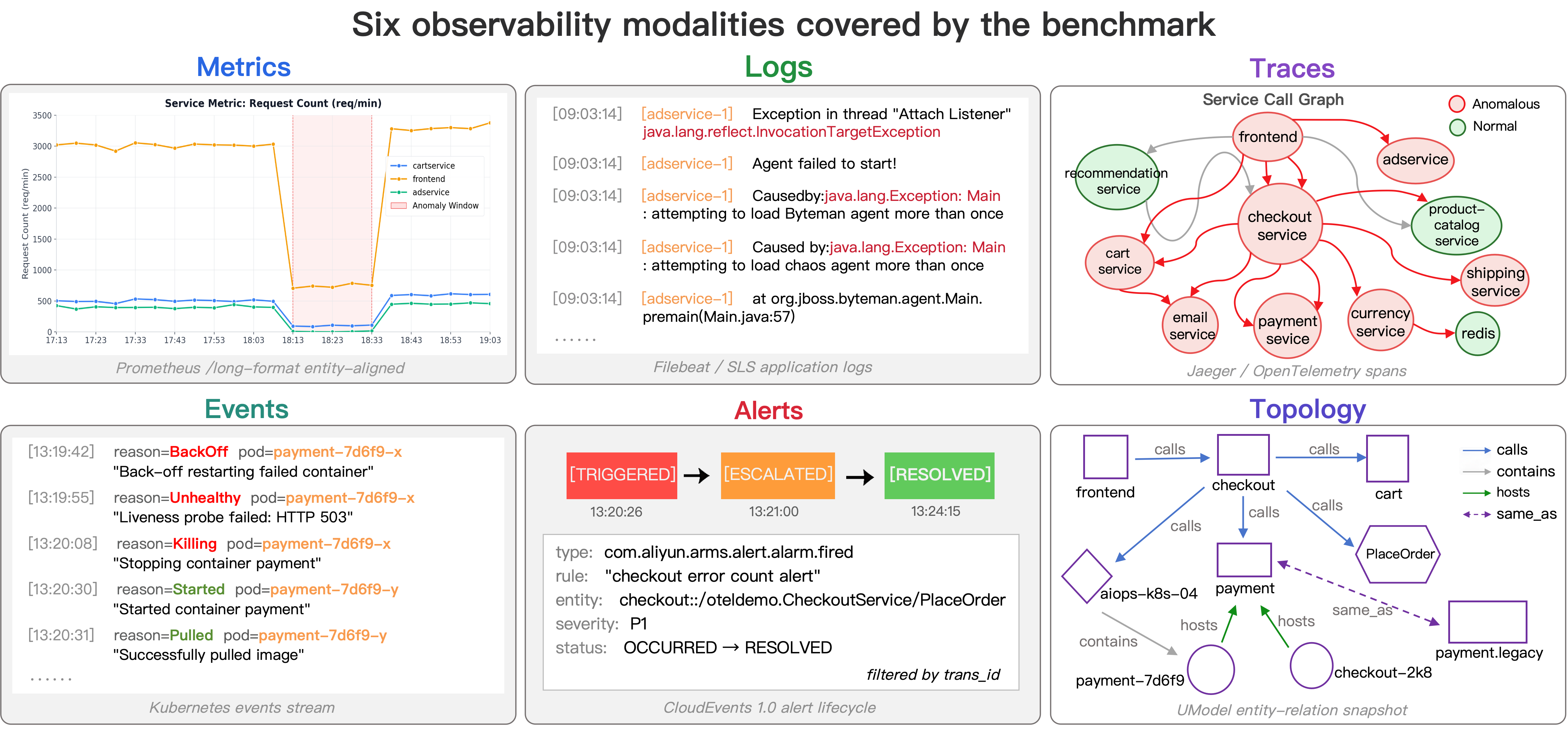}
  \caption{Multimodal observability data across the two datasets.
  The top three modalities (Metrics, Logs, Traces) constitute the
  base signal stack shared by \aiopsds and \rcads; the bottom
  three (Events, Alerts, Topology) are higher-order signals unique
  to \rcads. Anomalous behavior surfaces consistently across
  modalities via shared entity identifiers, letting an agent
  triangulate evidence along service / pod / entity dimensions.}
  \label{fig:modalities}
\end{figure*}

\subsection{Multimodal Coverage}
\label{sec:design:multimodal}

Diagnostic evidence is distributed across modalities with
complementary roles: metrics expose quantitative trends, logs
carry semantic error context, and traces record propagation along
the call topology. None is individually sufficient---similar
metric curves stem from disparate causes, log errors may never
be written before a container chokes, and JVM-internal anomalies
are invisible to cross-service spans. A high-quality benchmark
must therefore \emph{force} cross-modal fusion rather than allow
shortcuts through any single modality.

We therefore require \textbf{multimodal coverage}. \aiopsds
provides the three base modalities; \rcads further adds Events,
Alerts, and Topology (Fig.~\ref{fig:modalities}).

\subsection{Hierarchical Fault Coverage}
\label{sec:design:hierarchy}

Microservice faults differ systematically in observability
footprint, propagation scope, and reasoning difficulty depending
on \emph{where} and \emph{what} they are. Restricting a benchmark
to one entity layer or one fault family biases evaluation toward
methods tuned to that slice and obscures cross-layer
discrimination. We therefore require \textbf{hierarchical fault
coverage} along two orthogonal axes.

\paragraph{Entity layer} \emph{Service-level} faults affect every
pod under a service (e.g., network attacks, erroneous deploys);
\emph{pod-level} faults touch a single instance and demand
discrimination among replicas (e.g., pod failure / kill, I/O
pressure); \emph{node-level} faults span all services on the same
host (e.g., node CPU / memory / disk pressure).

\paragraph{Fault category} The benchmark covers typical
production fault families: \emph{resource} (CPU, memory, disk,
I/O), \emph{network} (latency, loss, DNS), \emph{runtime}
(JVM CPU / GC / exception / latency), \emph{middleware \&
database} (TiDB and Redis disturbances), and
\emph{application-logic} (erroneous deploy, traffic surge, rate
limiting, null-pointer). Both datasets are designed under this
principle.

\subsection{Reasoning-Process Label}
\label{sec:design:annot}

Traditional RCA benchmarks label only the final root-cause
component---adequate for statistical and causal-graph methods,
but insufficient under the LLM-agent regime, as it cannot
separate evidence-grounded reasoning from keyword luck. From a
fault perspective, a real failure is a propagation path along the
call topology, not a single component; from an agent perspective,
an LLM diagnoses step by step, not in one shot. A reasoning-process
label must therefore expose both the fault's propagation structure
and the per-step evidence the agent should consult along the way.

We therefore require \textbf{reasoning-process labels} that record,
beyond the final root cause, the propagation structure of the
fault and the evidence supporting each step of the diagnostic
trace.

\section{AIOps2025: 2025 CCF AIOps Challenge Dataset}
\label{sec:dataset_a}

\aiopsds is built for the 2025 CCF AIOps Challenge and contains
$400$ fault cases drawn from a single distribution. The input is an
open-ended natural-language anomaly description with a time
window---mimicking free-text alert triage; the agent must perform
data loading, anomaly detection, and cross-modal reasoning by
itself, and is scored not only on the predicted root-cause component
but also on the per-modality key evidence its reasoning trace covers.

\subsection{System and Fault Injection}
\label{sec:dataset_a:system}

\begin{figure}[t]
  \centering
  \includegraphics[width=\columnwidth]{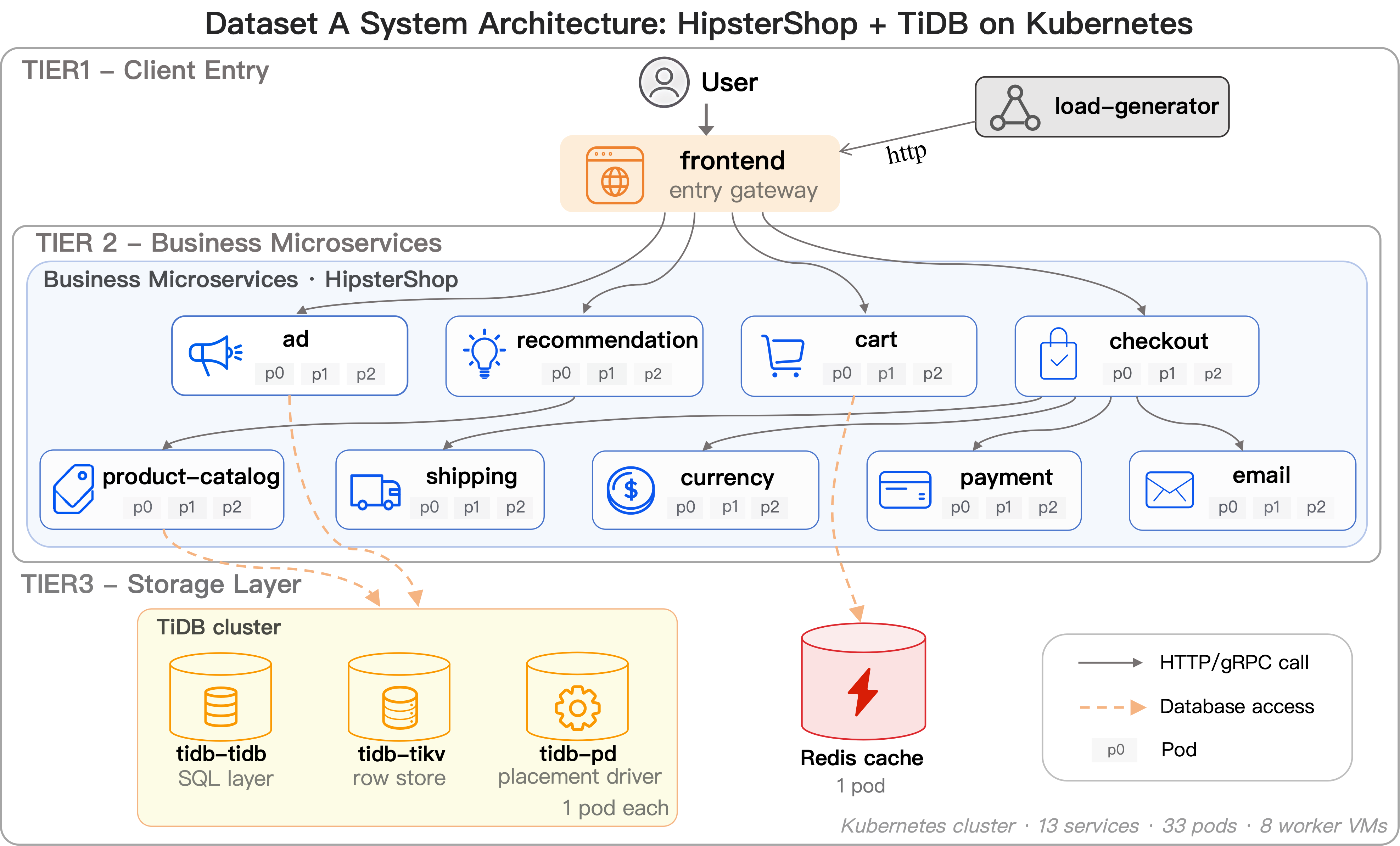}
  \caption{\aiopsds system architecture. Three tiers (client entry /
  business microservices / storage) on Kubernetes: $10$ HipsterShop
  services with $3$ pods each, three TiDB components, and a Redis
  cache, deployed across $8$ worker VMs.}
  \label{fig:dataset_a_arch}
\end{figure}

\paragraph{System} The underlying system spans three tiers
(Fig.~\ref{fig:dataset_a_arch}): the \emph{client entry} tier
(\emph{frontend} gateway plus a load-generator), the \emph{business
microservice} tier built on \textbf{HipsterShop} (Google Cloud's
open-source $10$-microservice e-commerce demo)~\cite{hipstershop},
and the \emph{storage} tier with \textbf{TiDB}~\cite{tidb} (PD, TiKV,
TiFlash) as the distributed SQL backend and a \textbf{Redis} cache.
The whole stack is orchestrated by Kubernetes across $8$ virtual
machines, totalling \textbf{$13$ services and $33$ pods} ($10 \times 3$
HipsterShop pods plus three TiDB components, $1$ pod each).
HipsterShop's call graph exercises synchronous HTTP/gRPC,
asynchronous messaging, and cross-language services.
\textbf{Differing from prior AIOps datasets that build on bare
HipsterShop and confine fault scope to the application
layer, \aiopsds also injects faults onto
the storage tier}---resource, I/O, and network disturbances on
TiDB components (PD, TiKV, TiFlash) and the Redis cache pod---so the
benchmark exercises diagnostic reasoning across the
application~$\to$~storage propagation path that production
incidents traverse but pure microservice demos systematically omit.

\paragraph{Fault injection} Faults are injected via
\textbf{Chaos-Mesh}~\cite{chaosmesh} at three hierarchical entity
levels in line with Section~\ref{sec:design:hierarchy}: service,
pod, and node. The dataset spans $9$ fault categories and $18$
fault types covering resource, network, runtime, application-logic,
and infrastructure failures; Table~\ref{tab:a-faults} summarizes
the categories with per-type case counts. Per-category injection
parameters (durations, magnitudes, target-instance selection) are
randomized within fixed ranges, released alongside the dataset.

\begin{table}[t]
\footnotesize
\centering
\caption{\aiopsds fault taxonomy: $9$ categories $\times$ $18$
fault types, with per-type case counts.}
\label{tab:a-faults}
\setlength{\tabcolsep}{3pt}
\begin{tabular}{p{0.18\columnwidth}p{0.20\columnwidth}p{0.40\columnwidth}r}
\toprule
Category & Fault type & Description & \#Cases \\
\midrule
\multirow{3}{*}{node fault}
  & node cpu stress    & Node-level CPU saturation                   & 23 \\
  & node memory stress & Node-level memory saturation                & 38 \\
  & node disk fill     & Node-level disk exhaustion                  & 21 \\
\midrule
\multirow{3}{*}{network attack}
  & network delay      & Inter-service latency injection             & 25 \\
  & network loss       & Packet loss between services                & 21 \\
  & network corrupt    & Packet corruption between services          & 27 \\
\midrule
\multirow{2}{*}{pod fault}
  & pod failure        & Whole-service or single-pod down            & 45 \\
  & pod kill           & One-shot pod restart                        & 15 \\
\midrule
\multirow{4}{*}{jvm fault}
  & jvm cpu            & JVM CPU spike                               & 13 \\
  & jvm gc             & Stop-the-world GC pause                     & 14 \\
  & jvm exception      & Java exception throwing                     & 13 \\
  & jvm latency        & JVM-internal latency injection              & 15 \\
\midrule
\multirow{2}{*}{stress test}
  & cpu stress         & Service-/pod-level CPU stress               & 22 \\
  & memory stress      & Service-/pod-level memory stress            & 20 \\
\midrule
io fault         & io fault             & Disk I/O delay or error                     & 28 \\
\midrule
erroneous change & code error           & Buggy image deployed to a service           & 21 \\
\midrule
dns fault        & dns error            & DNS resolution failure                      & 21 \\
\midrule
misconfiguration & target port misconfig & Service targetPort mis-binding    & 18 \\
\midrule
\textbf{Total}   & \textbf{18 types}    &                                             & \textbf{400} \\
\bottomrule
\end{tabular}
\end{table}

\subsection{Multimodal Data}
\label{sec:dataset_a:data}

Each case ships its full Metrics, Logs, and Traces over the
anomaly window. Metrics come from Prometheus~\cite{prometheus}
(service/pod APM) plus node and TiDB-cluster infrastructure
feeds; Logs from Filebeat; Traces from Jaeger~\cite{jaeger}. The
three modalities share service and pod names,
anchoring an event to consistent identifiers across modalities;
samples are shown in Fig.~\ref{fig:modalities} (top). The dataset
comprises \textbf{$2{,}835$ Parquet files, $\approx 269$\,M rows,
and $11.9$\,GB} over $18$ calendar days, partitioned by phase /
calendar day / modality and sliced hourly so participants can
load only the relevant range without touching the full corpus.

\subsection{Groundtruth Labeling}
\label{sec:dataset_a:io}

To ensure label reliability, each case is labelled through a
three-stage \emph{algorithm-plus-multi-expert} pipeline. \emph{(i)}
Multimodal anomaly detectors extract candidate evidence points
from raw Metrics, Logs, and Traces; \emph{(ii)} three SRE experts
independently confirm or revise the candidates per case without
seeing each other's outputs; \emph{(iii)} a senior SRE expert
adjudicates remaining disagreements. Every fault scenario is also
\emph{re-injected multiple times} before admission, so published
labels reflect what a domain expert~\cite{sre-book, onlinemoira}
\emph{should observe} in the data, not what the injection command
did.

Each case's ground truth (Fig.~\ref{fig:gt-a}) pairs the fault
metadata---a (category, type, instance) triple plus localization
fields whose semantics depend on the (instance type, fault
category) combination, since a network attack centers on a
source--destination service pair while a node fault centers on a
node---with three reasoning-trace labels: a list of per-modality
key observations and a list of must-hit metric names the agent's
trace should cover (driving the Explainability score), and a list
of semantically equivalent phrasings of the fault type (driving
the Type Accuracy score, Section~\ref{sec:dataset_a:eval}).

\begin{figure}[t]
\begin{gtbox}
\begin{lstlisting}[language=gtjson]
{
  "uuid": "345fbe93-80",
  "fault_category": "stress test",
  "fault_type":     "cpu stress",
  "instance_type":  "service",
  "service":  "emailservice",
  "instance": "emailservice",
  "start_time": "2025-06-05T16:10:02Z",
  "end_time":   "2025-06-05T16:31:02Z",

  "key_observations": [
    {"type": "metric", "keyword": ["pod_processes"]},
    {"type": "metric", "keyword": ["rrt", "rrt_max"]},
    {"type": "metric", "keyword": ["pod_cpu_usage"]}
  ],
  "key_metrics": ["pod_cpu_usage"],
  "fault_description": [
    "high CPU utilization", "CPU saturation",
    "CPU overload", "CPU usage spike", "cpu stress"
  ]
}
\end{lstlisting}
\end{gtbox}
\caption{\aiopsds ground-truth example: a service-level
\emph{cpu stress} case on \emph{emailservice}. Per-modality
\emph{key\_observations} and \emph{key\_metrics} drive the
Explainability score; \emph{fault\_description} drives the Type
Accuracy score.}
\label{fig:gt-a}
\end{figure}

\subsection{Evaluation Metric}
\label{sec:dataset_a:eval}

\paragraph{Design rationale} The protocol is organized along the
three pillars of Section~\ref{sec:intro} and weighted
$0.40 / 0.40 / 0.20$: \emph{Localization} and \emph{Identification} ($80\%$)
form the actionable conclusion, and \emph{Reason} ($20\%$) is split
into two interlocking process dimensions that constrain each other
to prevent reward hacking on the reasoning trace.

\paragraph{Localization via Location Accuracy (LA)} LA is the
fraction of cases whose predicted component string-matches the
ground truth exactly:
\begin{equation}
\mathrm{LA} = \frac{L_c}{L_t}.
\end{equation}
Strict matching distinguishes ``right service'' from ``right
instance'' (e.g., \emph{emailservice} $\neq$
\emph{emailservice-0}) and prevents fuzzy-matching score
inflation. Two fault families need targeted relaxations: for
network attacks, LA accepts a hit on either source or destination
since evidence is prominent at both endpoints; for pod-level
faults, LA requires the exact pod identifier rather than just the
service.

\paragraph{Identification via Type Accuracy (TA)} TA scores
whether the agent's reason field faithfully reflects the true
fault type via two-step matching: keyword detection against the
ground-truth fault descriptions (any hit yields full credit),
with embedding-similarity fallback for unhit cases. The reason
field is truncated to its first $20$ words to suppress keyword
stuffing.

\paragraph{Reason via Explainability and Efficiency} The Reason
pillar is captured by two complementary process metrics. The core
metric is \textbf{Explainability} ($0.10$), measuring evidence-point
coverage:
\begin{equation}
\mathrm{Explainability} = \frac{E_m}{E_t},
\end{equation}
where $E_t$ is the total expected evidence points (the union of
the GT's key metrics and key observations) and $E_m$ the count
hit in any of the agent's trace observations under strict
metric-name, log-keyword, or trace-node matching; each
observation contributes only its first $20$ characters, again to
suppress keyword stuffing. \textbf{Efficiency} ($0.10$)
counter-balances Explainability by penalizing overly long traces
on LA-correct cases:
\begin{equation}
\mathrm{Efficiency} = \min\!\Big(1.0,\ \exp\!\big(-\tfrac{APL - 5}{5}\big)\Big),
\end{equation}
where $APL$ is the mean \emph{reasoning\_trace} length over
LA-correct cases. Restricting to LA-correct prevents wrong-answer
short-path gaming; together, the two metrics reward traces that
are evidence-grounded \emph{and} concise.

\paragraph{Final score} The four dimensions aggregate into a
$0$--$100$ score:
\begin{multline}
\mathrm{Final}_A = (0.4\,\mathrm{LA} + 0.4\,\mathrm{TA} + 0.1\,\mathrm{Exp.} + 0.1\,\mathrm{Eff.}) \times 100.
\end{multline}

\subsection{Empirical Analysis}
\label{sec:dataset_a:props}

\begin{table}[t]
\footnotesize
\centering
\caption{\aiopsds statistical properties.}
\label{tab:a-props}
\setlength{\tabcolsep}{4pt}
\begin{tabular}{p{0.32\columnwidth}p{0.58\columnwidth}}
\toprule
Metric & Value \\
\midrule
Injection-level distribution      & service $195$ / pod $123$ / node $82$ \\
Cross-modal necessity             & $\geq\!2$ modalities $250 / 400$ ($62.5\%$); all $3$ $124 / 400$ ($31\%$) \\
Per-modality marginal need        & metric $92.8\%$, log $56.2\%$, trace $43.0\%$ \\
Total evidence entries            & $1{,}878$ (mean $4.7$ per case); modality split metric $50.8\%$, log $40.0\%$, trace $9.2\%$ \\
\bottomrule
\end{tabular}
\end{table}

\paragraph{Difficulty distribution} Difficulty is shaped by the
coupling between fault category and injection level. The $400$
cases distribute across the three injection levels as
\textbf{service $195$ / pod $123$ / node $82$}
(Table~\ref{tab:a-props}). Most categories tie to one
``natural'' level---node fault at node, network attack at
service---while four categories (pod fault, jvm fault, stress
test, and dns fault) span both service and pod, so the three
injection levels expose distinct diagnostic patterns rather than
identical cases at different scopes.

\paragraph{Cross-modal reasoning necessity} Co-occurrence
signatures over the per-case key observations show that
\textbf{$62.5\%$ of cases require $\geq\!2$ modalities and $31\%$
require all three} (Table~\ref{tab:a-props}); per-modality
marginal necessity is graded (metric $92.8\%$, log $56.2\%$, trace
$43.0\%$), providing data-layer evidence for the
multimodal coverage principle of
Section~\ref{sec:design:multimodal}.

\paragraph{Large-scale real-world validation} The 2025 CCF AIOps
Challenge---the first edition of this long-running venue to move
into the LLM-agent regime---uses \aiopsds under the
four-dimensional protocol of Section~\ref{sec:dataset_a:eval}. It
attracted \textbf{$561$ teams and $1{,}068$ contestants} from
leading universities and industry, demonstrating the protocol's
tractability for end-to-end agent submissions at production scale.

\section{RCA100: 2025 Alibaba Tianchi AIOps Dataset}
\label{sec:dataset_b}

\rcads is built for the Tianchi 2025 AIOps Track
(Section~\ref{sec:dataset_b:props}) and instantiates causal-chain
coverage. It comprises \textbf{$103$ fault events} injected via
chaos drills on the OpenTelemetry Demo Store deployed on Alibaba
Cloud ACK, with \textbf{six modalities}---Metrics, Logs, Traces,
Events, Alerts, and Topology---totalling $\approx 3.4$\,GB and
released under CC~BY-NC-SA~4.0.

\subsection{System and Fault Injection}
\label{sec:dataset_b:system}

\begin{figure}[t]
  \centering
  \includegraphics[width=\columnwidth]{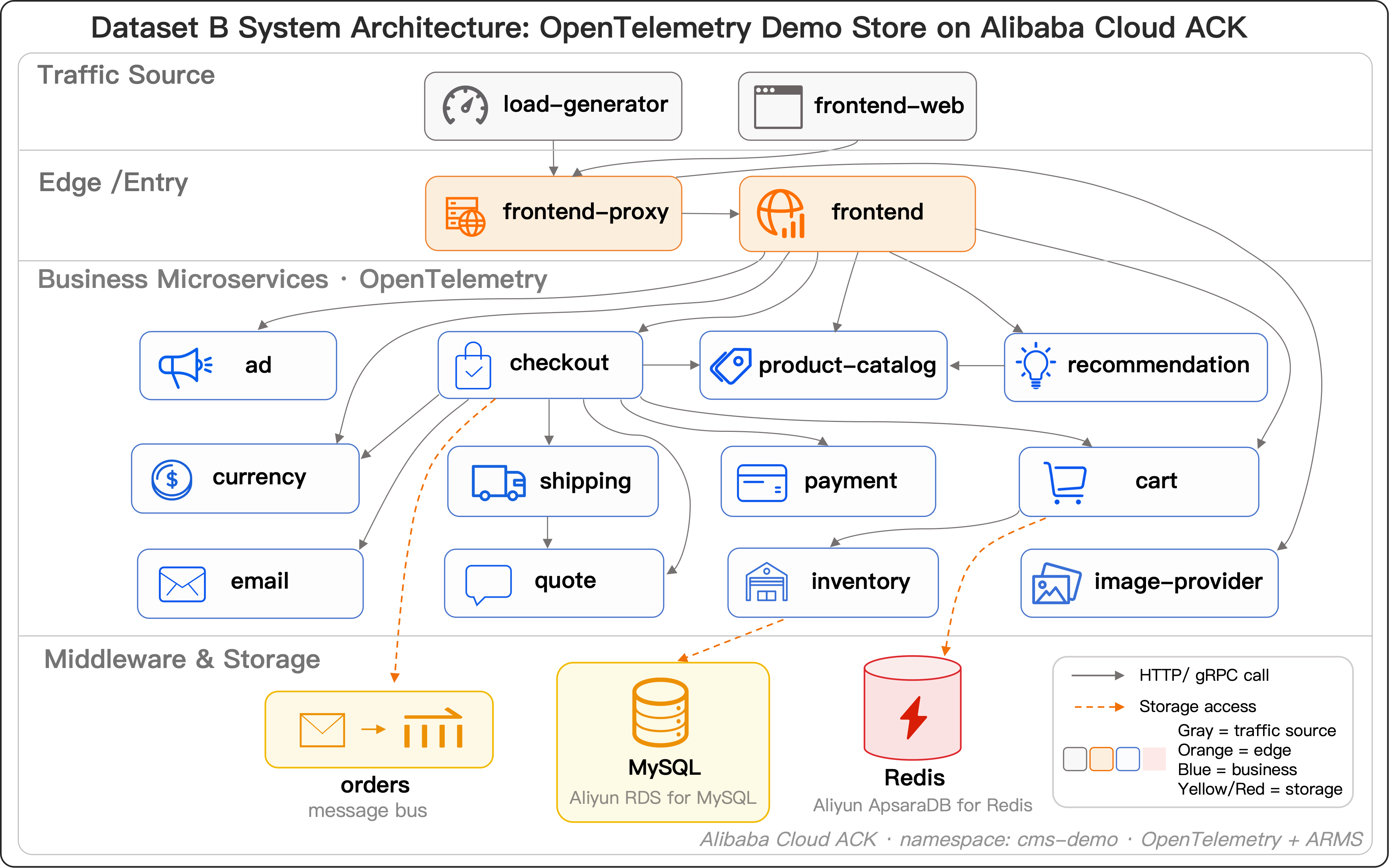}
  \caption{\rcads system architecture. The OpenTelemetry Demo
  Store on Alibaba Cloud ACK: a polyglot microservice e-commerce
  application entered through \emph{frontend} $\to$
  \emph{frontend-proxy}, backed by Aliyun RDS for MySQL,
  ApsaraDB for Redis, and an order message bus, with traces
  collected through a hybrid OpenTelemetry $+$ ARMS Agent pipeline.}
  \label{fig:dataset_b_arch}
\end{figure}

\paragraph{System} The underlying system is the open-source
\textbf{OpenTelemetry Demo Store}~\cite{opentelemetry}, a polyglot
e-commerce application of roughly a dozen microservices (Java,
C++, Rust, Python) deployed on an \textbf{Alibaba Cloud
ACK} cluster with managed RDS, Redis, and message-bus backends
(Fig.~\ref{fig:dataset_b_arch}). Telemetry is collected through a
hybrid \textbf{OpenTelemetry $+$ ARMS Agent} pipeline, with the
ARMS Agent surfacing JVM-internal state and node-level
operational signals not captured by stock OTel.
\textbf{Aliyun UModel} provides the unified entity-type system
that resolves the same logical entity---named differently in APM,
K8s, and cloud-resource views---into a single entity ID,
enabling chain-step comparability across domains.

\paragraph{Fault injection} Faults are injected via
\textbf{Chaos Drills}, producing $103$ events that span $28$
root-cause types aggregated into six semantic groups
(Table~\ref{tab:b-faults}). Each case starts from a single alert
event as the diagnostic entry point: $90 / 103$ cases carry a
single alert entity (typically at the apm.operation level), while
the remaining $13$ are kept as composite scenarios with no alert
entity to avoid biasing the benchmark toward cases that start
from a single entity. Answers are distributed only via an
independent answer-key package, never exposed in the public
task contract, ensuring blind-evaluation fairness.

\begin{table}[t]
\footnotesize
\centering
\caption{\rcads fault taxonomy: $28$ root-cause types
aggregated into $6$ semantic groups, with per-type case counts.}
\label{tab:b-faults}
\setlength{\tabcolsep}{4pt}
\begin{tabular}{p{0.27\columnwidth}p{0.50\columnwidth}r}
\toprule
Group & Root-cause type & \#Cases \\
\midrule
\multirow{7}{*}{\shortstack[l]{Application logic\\($38$)}}
  & httpError5xx            & 11 \\
  & rateLimiting            & 11 \\
  & trafficSurge            &  6 \\
  & nullPointerException    &  4 \\
  & trafficHotspot          &  3 \\
  & loadBalancerFailure     &  2 \\
  & codeDefect              &  1 \\
\midrule
\multirow{3}{*}{\shortstack[l]{JVM runtime\\($16$)}}
  & memoryPressure          & 12 \\
  & threadExhaustion        &  2 \\
  & fullGC                  &  2 \\
\midrule
\multirow{3}{*}{\shortstack[l]{Cloud resource\\($14$)}}
  & nodeCpuHigh             & 12 \\
  & nodeDown                &  1 \\
  & nodeMemoryOOM           &  1 \\
\midrule
\multirow{5}{*}{\shortstack[l]{Middleware \& DB\\($13$)}}
  & slowSQL                 &  5 \\
  & redisUnavailable        &  4 \\
  & dbNetworkLatency        &  2 \\
  & messageQueueBacklog     &  1 \\
  & cacheBreakdown          &  1 \\
\midrule
\multirow{7}{*}{\shortstack[l]{K8s lifecycle\\($12$)}}
  & replicaScaleDown        &  6 \\
  & resourceLimitMisconfig  &  1 \\
  & podCrashLoop            &  1 \\
  & podPendingUnschedulable &  1 \\
  & podRestartFlapping      &  1 \\
  & networkPolicyIsolation  &  1 \\
  & dnsResolutionFailure    &  1 \\
\midrule
\multirow{3}{*}{\shortstack[l]{Resource \& perf.\\($10$)}}
  & cpuFullLoad             &  8 \\
  & cpuDeadLoop             &  1 \\
  & diskIOHigh              &  1 \\
\midrule
\textbf{Total} & \textbf{$28$ types ($6$ groups)} & \textbf{$103$} \\
\bottomrule
\end{tabular}
\end{table}

\subsection{Multimodal Data}
\label{sec:dataset_b:data}

Each task ships as a self-contained diagnostic slice of $7$ files:
$5$ Parquets (one each for metrics, logs, traces, events, and
alerts) plus the agent-facing task contract and an
entity-relation topology snapshot. Metrics use a long-format
entity-aligned schema in which every row carries an entity ID
resolving to a UModel topology entity; Logs and Traces follow
the SLS application-log and OpenTelemetry span schemas. Modality
samples are in Fig.~\ref{fig:modalities}
(Section~\ref{sec:design:multimodal}).

\paragraph{Higher-order modalities} The three signals beyond the
base $M+L+T$ stack play distinct semantic roles. \emph{Events}
stream K8s lifecycle signals (pod restart, scheduling failure,
backoff) that are visible only in this modality and indispensable
for pod- and node-layer faults. \emph{Alerts} provide the
entry-alert lifecycle ($\sim\!30$ events per case, covering
trigger, escalation, and recovery). \emph{Topology} is a per-task
UModel entity-relation snapshot at the alert moment, serving as
the graph skeleton for cross-modal reasoning.

\paragraph{Scale and integrity} In total \rcads comprises
\textbf{$721$ files, $\approx 116$\,M rows, $3.4$\,GB}. We verified
full-corpus reference integrity: every cross-modality reference
resolves into the corresponding task's topology at $100\%$ with no
dangling edges, and GT root-cause entities match the topology at
$98.06\%$ ($101 / 103$).

\begin{table*}[!t]
\centering
\caption{Design comparison of \aiopsds and \rcads.}
\label{tab:dataset-compare}
\small
\begin{tabular}{p{0.18\textwidth}p{0.38\textwidth}p{0.38\textwidth}}
\toprule
Dimension & \aiopsds & \rcads \\
\midrule
Core target  & Reasoning-process explainability & Causal-chain reasoning correctness \\
Scale        & $400$ cases & $103$ cases \\
System       & HipsterShop on self-hosted K8s ($\sim\!13$ services) & OTel Demo Store on Aliyun ACK ($\sim\!$ a dozen services) \\
Modalities   & $3$ (Metrics / Logs / Traces) & $6$ ($+$Events / Alerts / Topology) \\
Data layout  & Per day $\times$ hour; shared across cases & Per-task self-contained ($7$ files per task) \\
Data scale   & $2{,}835$ files / $269$\,M rows / $11.9$\,GB & $721$ files / $116$\,M rows / $3.4$\,GB \\
Input form   & Open NL anomaly description $+$ time window & Structured alert event in NL prompt ($90/103$ with alert entity) \\
Topology     & Implicit (must be inferred) & Explicit entity graph (per-task snapshot) \\
Taxonomy     & $9$ categories $\times$ $18$ types $\times$ $3$ levels & $28$ root-cause types $\to$ $6$ semantic groups \\
Causal chain & None (single-step localization) & $3$--$4$ steps explicit ($91$ / $12$) \\
Evidence    & Per-modality key-evidence list & Anchored to chain steps (checkpoints) \\
Evaluation   & LA / TA / Exp.\ / Eff.\ at $0.40/0.40/0.10/0.10$ & Entity / Fault / Process at $0.40/0.30/0.30$ \\
\bottomrule
\end{tabular}
\end{table*}

\subsection{Groundtruth Labeling}
\label{sec:dataset_b:contract}

The four-layer labels are produced under the same
algorithm-plus-multi-expert pipeline as \aiopsds
(Section~\ref{sec:dataset_a:io}): multimodal detectors propose
candidate chain steps and observability checkpoints, three SRE
experts independently revise them per case, and a senior expert
adjudicates disagreements. Every chaos drill is also re-injected
multiple times before admission.

Each case's ground-truth file (Fig.~\ref{fig:gt-b}) carries four
layers: an expected fault type drawn from the $28$-class
taxonomy, a target-entity list pinning the root cause to UModel
entity IDs, a reasoning-step array encoding the causal chain as
cause~$\to$~propagation~$\to$~impact ($91$ three-step and $12$
four-step chains), and per-step observability checkpoints. The
$661$ checkpoints ($\approx 6.42$ per case) span metric, trace,
event, and alert sources, and $99.5\%$ carry a
$\langle$comparator, value, unit$\rangle$ numeric constraint, so
the agent must \emph{match the numeric condition}, not merely
mention the metric. Common signals include request count ($204$
checkpoints), average request latency ($178$), and error count
($97$).

\begin{figure}[t]
\begin{gtbox}
\begin{lstlisting}[language=gtjson]
{
  "case_id": "F014-httpError5xx.tbdh9alum56k...",
  "outcome": {
    "expected_fault_id": "F014-httpError5xx",
    "target_entities": [
      {"entity_name": "payment", "entity_type": "apm.service"}
    ]
  },
  "reasoning": { "steps": [
    {"step": 1, "step_type": "cause",
     "target": "payment",
     "observability": [
       {"signal": "error_count",
        "expected": {"comparator": ">=", "value": 8829}},
       {"signal": "error_rate",
        "expected": {"comparator": ">=", "value": 0.50}}
     ]},
    {"step": 2, "step_type": "propagation",
     "target": "checkout",
     "observability": [
       {"signal": "error_count",
        "expected": {"comparator": ">=", "value": 8828}},
       {"signal": "error_rate",
        "expected": {"comparator": ">=", "value": 0.50}}
     ]},
    {"step": 3, "step_type": "impact",
     "target": "checkout::PlaceOrder",
     "observability": [
       {"signal": "error_count",
        "expected": {"comparator": ">=", "value": 8828}}
     ]}
  ]}
}
\end{lstlisting}
\end{gtbox}
\caption{\rcads ground-truth example: an
httpError5xx case propagating along
\emph{payment}~$\rightarrow$~\emph{checkout}~$\rightarrow$~\emph{PlaceOrder}.
Each chain step carries a typed role and numeric observability
checkpoints the agent must match.}
\label{fig:gt-b}
\end{figure}

\subsection{Evaluation Metric}
\label{sec:dataset_b:eval}

\paragraph{Design rationale} The protocol is organized along the
three pillars of Section~\ref{sec:intro} and weighted
$0.40 / 0.30 / 0.30$: Entity Localization leads ($0.40$) since
the located entity is the actuation point of any repair; Fault
Identification and Reasoning Process get equal $0.30$, since
naming the right fault type and walking the right chain
contribute symmetrically to diagnostic credibility. The first
two dimensions ($70\%$) are quantified deterministically via
UModel topology distance, with no LLM-as-judge component in the
bulk of the protocol.

\paragraph{Localization via Entity Localization} A UModel
entity-ID match with partial credit on topologically adjacent
entities ($0.40$).

\paragraph{Identification via Fault Identification} A
fine-grained $28$-class match against the expected fault type
($0.30$).

\paragraph{Reason via Reasoning Process} The Reason pillar is
captured by a process score ($0.30$) jointly measured by
causal-chain node match rate and observability-checkpoint hit
rate, so an agent must walk the right chain \emph{and} cite the
right evidence at each step.

\paragraph{Final score} The three dimensions aggregate into a
$0$--$100$ score:
\begin{multline}
\mathrm{Final}_B = (0.4\,\mathrm{Entity} + 0.3\,\mathrm{Fault} + 0.3\,\mathrm{Process}) \times 100.
\end{multline}

\subsection{Empirical Analysis}
\label{sec:dataset_b:props}

\begin{table}[t]
\footnotesize
\centering
\caption{\rcads statistical properties.}
\label{tab:b-props}
\setlength{\tabcolsep}{4pt}
\begin{tabular}{p{0.32\columnwidth}p{0.58\columnwidth}}
\toprule
Metric & Value \\
\midrule
Root-cause-level distribution    & apm.svc $83$ / k8s.node $17$ / k8s.pod $3$ \\
Fault-type concentration         & top-$4$ types $44.7\%$; remaining $24$ form a long tail \\
Multi-entity chains              & $95 / 103$ ($92.2\%$) traverse $\geq\!2$ entity kinds \\
Cross-domain chains              & $20 / 103$ ($19.4\%$) cross APM~$\leftrightarrow$~K8s \\
Avg.\ modalities touched         & $4.02 / 6$ per case \\
Step-type counts                 & cause $103$ / propagation $115$ / impact $103$ \\
\bottomrule
\end{tabular}
\end{table}

\paragraph{Difficulty distribution} Difficulty is
multidimensional. The $28$ root-cause types follow a long-tail
distribution: the top four (nodeCpuHigh, memoryPressure,
httpError5xx, and rateLimiting) account for $44.7\%$ of cases
while the remaining $24$ form the long tail. Root causes sit at
the APM service level in $83$ cases, the K8s node level in $17$,
and the K8s pod level in $3$ (Table~\ref{tab:b-props}); the
$20/103$ ($19.4\%$) cases whose chain crosses the
APM~$\leftrightarrow$~K8s domain boundary form the hardest subset
for cross-domain attribution.

\paragraph{Cross-modal reasoning necessity} Although $98.2\%$ of
the $661$ checkpoints sit on metric, the other modalities carry
\emph{structural} rather than counted evidence: $92.2\%$ of
chains traverse $\geq\!2$ entity kinds (relying on Topology's
cross-domain alias edges for normalization), and the same
$19.4\%$ of cases with K8s-layer root causes depend on lifecycle
evidence visible only in Events. On average each case touches
$4.02$ of the $6$ modalities.

\paragraph{Large-scale real-world validation} The Tianchi 2025
AIOps Track---a track of Alibaba Cloud's \emph{2025 AI-Native
Programming Challenge}---uses \rcads under the
three-dimensional protocol of Section~\ref{sec:dataset_b:eval}.
It attracted \textbf{$5{,}532$ teams} of cloud-platform
practitioners and academic researchers, validating the
causal-chain protocol at production scale.

\subsection{Comparison of the Two Datasets}
\label{sec:dataset_b:compare}

Table~\ref{tab:dataset-compare} contrasts the two datasets along
$12$ dimensions. They share the same essence---multi-layer
microservice telemetry evaluated across modalities---and differ in
how they expose the reasoning process: \aiopsds labels evidence as
a flat per-modality list (\emph{key-evidence coverage}), while
\rcads labels it as a typed causal chain
(\emph{causal-chain coverage}). Both raise the explainability floor
of agent diagnosis. As supervision signals for agent training the
two are complementary: key-evidence labels are cheaper to obtain
at scale, while causal-chain labels are more expensive but more
informative, since they expose how evidence chains into a
propagation explanation. Future microservice RCA datasets should
invest in causal-chain labelling whenever the annotation budget
allows.

\section{Discussion}
\label{sec:lessons}

\subsection{Fault Injections $\neq$ Fault Cases}
\label{sec:lessons:validity}

A common misconception in chaos-driven benchmark construction is
to equate ``a fault injection'' with ``a usable fault case.'' The
loss between the two is substantial, in three recurring patterns:
\emph{(i) absorption by fault tolerance}---retries, degradation,
or autoscaling silently mask the fault; \emph{(ii) faint
footprints}---an alert fires but the trace is too thin for even a
domain expert to reverse-engineer a complete causal chain;
\emph{(iii) divergence from intent}---e.g., CPU pressure
unexpectedly triggering an OOM kill, so the GT label and actual
behavior fall out of sync. \aiopsds's $400$ and \rcads's $103$
cases are accordingly drawn from substantially larger injection
pools, admitting only those with a complete, expert-traceable
observability footprint. \emph{Benchmark difficulty is not the
difficulty of injecting a fault, but the difficulty of diagnosing
one from its observability footprint.}

\subsection{Data Cleaning for Downstream Reuse}
\label{sec:lessons:cleaning}

Raw observability data is voluminous, inconsistently named, and
laced with irregular cross-modal references; released as is, most
participant effort would go into preprocessing rather than
diagnosis. We therefore cleaned both datasets along three axes:
\emph{cross-modal entity alignment} (service/pod names in
\aiopsds, UModel entity IDs in \rcads),
\emph{redundant-slice trimming} (per-task self-contained files in
\rcads; per domain/day/hour partitioning in \aiopsds), and
\emph{field standardization} (timestamps, nulls, modality
identifiers, Parquet types).

This is not merely a benchmark convenience. Production telemetry
exposes the same heterogeneity at greater scale---fragmented across
time-series databases (Prometheus, InfluxDB), log backends
(Elasticsearch), tracing stores, and per-vendor APIs. An agent
dropped onto raw production data must adapt to each surface before
any reasoning can begin, inflating engineering cost and amplifying
hallucination. A more workable direction is to push unification
into a dedicated tool layer---e.g., agent-ready observability data
modeling~\cite{umodel}---so the agent queries one normalized
interface rather than negotiating $N$ backends. Our cleaning
mirrors that pattern at benchmark scale.

\subsection{Limitations}
\label{sec:disc:limitations}

We acknowledge four limitations.
\emph{(i) Scale gap to production.} Both datasets are built on
open-source demos of around a dozen services; production systems
can run hundreds to thousands across complex middleware and
cross-region deployments, and transfer to that scale remains to
be verified.
\emph{(ii) Boundary of fault realism.} Chaos-injected faults are
relatively clean (single root cause, controlled parameters),
whereas production failures often present as concurrent causes,
intermittent flares, or long-accumulated degradation.
\emph{(iii) Boundary of label coverage.} Even with multi-expert
labelling, valid but non-mainstream diagnostic paths can be
omitted; agents that reach the correct root cause via an
unlabelled trace are under-credited.
\emph{(iv) Insufficient coverage of efficiency.} \aiopsds's
Efficiency metric uses reasoning-trace length as a
proxy and does not capture end-to-end latency, token consumption,
or tool-call cost.

\subsection{Beyond RCA: Extended Uses of the Datasets}
\label{sec:lessons:beyond}

Multimodal completeness and fine-grained expert labels make both
datasets reusable beyond microservice RCA: the time-aligned
Metrics+Logs+Traces support multimodal time-series anomaly
detection; \rcads's causal-chain labels and entity-relation
graphs supply ground truth for causal discovery and graph neural
methods; \aiopsds's per-modality key-evidence labels are rare
material for studying expert--agent reasoning-path divergence
and agent self-evaluation; and the real-failure traces support
training fault-synthesis or fault-injection models.

More broadly, the underlying paradigm is not specific to
microservices: any task that rests an answer on evidence and
admits structural labelling of that evidence---medical diagnosis,
legal reasoning, scientific discovery---can adopt the same
key-evidence and causal-chain coverage forms, turning ``does the
agent's reasoning rest on the right evidence'' into a computable
scoring signal.

\section{Conclusion}
\label{sec:conclusion}

We presented two complementary microservice AIOps benchmarks
that operationalize \emph{reasoning-process evaluation}: $400$
HipsterShop cases anchor the diagnosis to per-modality key
evidence, and $103$ OpenTelemetry Demo Store cases anchor it to
typed causal chains over a UModel topology. Both have been
stress-tested at scale---$6{,}093$ teams across the 2025 CCF
AIOps Challenge and the Tianchi AIOps Track---and are released
with their scoring code. We hope the datasets shift agentic
diagnosis benchmarking from final-answer matching toward
evidence-grounded reasoning, in microservice RCA and in adjacent
evidence-driven agent tasks.


\bibliographystyle{ACM-Reference-Format}
\bibliography{references}

\end{document}